\documentclass[letter,twocolumn]{jpsj3}

\usepackage{amsmath}
\usepackage{amsfonts}
\usepackage{amssymb}
\usepackage{txfonts}
\usepackage{bm}
\usepackage{tabularx}
\usepackage{graphicx,color}
\usepackage{cite}

\def\Vec#1{\bm{#1}}
\def\Hc2{H_\mathrm{c2}}

\def\DTO{{\rm Dy_2Ti_2O_7}}

\author{
	Shunichiro \textsc{Kittaka}$^1$\thanks{kittaka@issp.u-tokyo.ac.jp}, 
	Shota \textsc{Nakamura}$^{1}$\thanks{Present address: Department of Engineering Physics, Electronics and Mechanics, Graduate School of Engineering, Nagoya Institute of Technology, Nagoya 466-8555, Japan}, 
	Hiroaki \textsc{Kadowaki}$^2$,\\
	Hiroshi \textsc{Takatsu}$^3$, and
	Toshiro \textsc{Sakakibara}$^{1}$ 
}

\inst{$^{1}$Institute for Solid State Physics, University of Tokyo, Kashiwa, Chiba 277-8581, Japan\\
      $^{2}$Department of Physics, Tokyo Metropolitan University, Hachioji-shi, Tokyo 192-0397, Japan\\
      $^{3}$Department of Energy and Hydrocarbon Chemistry, Graduate School of Engineering, Kyoto University, Kyoto 615-8510, Japan\\
}

\begin{document}

\title{Field-rotational Magnetocaloric Effect:\\ A New Experimental Technique for High-resolution Measurement\\ of the Anisotropic Magnetic Entropy}

\date{\today}

\abst{
We developed a new technique for measuring the thermodynamic entropy as a function of the magnetic field angle.
This technique enables high-resolution angle-resolved measurements of the entropy in an unprecedentedly short measuring time.
When the magnetic field is rotated under adiabatic conditions,
the sample temperature changes owing to the field-angle variation of its entropy,
which is referred to as the rotational magnetocaloric effect.
By investigating this effect along with the specific heat, the field-angle dependence of the entropy can be determined.
To demonstrate this technique, we chose the spin-ice compound Dy$_2$Ti$_2$O$_7$ as a benchmark 
and showed good agreement between the measured and theoretical entropies as a function of the field angle.
This development provides a new approach to studying condensed-matter physics, in which multiple degrees of freedom play an important role.
}

\maketitle
The entropy $S$ is one of the most fundamental thermodynamic quantities and is related to the number of microscopic configurations of the system.
It is very useful for understanding the ground state of a material that has multiple degrees of freedom. 
The entropy is usually evaluated from the temperature dependence of the specific heat $C$ as $S(T)=\int_0^T (C/T)dT$.
However, investigation of its field dependence, \textit{i.e.}, $S(H)$, by this method requires 
a number of measurements of $C(T)$ at various $H$ and accordingly requires an enormous amount of time to obtain high-resolution data.
For this purpose, measurement of the magnetocaloric effect (MCE), a temperature change in response to an adiabatic change in the external magnetic field, 
is more suitable; $S(H)$ can be directly evaluated as $S(H)-S(H=0)=-\int_0^H C/T (dT/dH) dH$.
This MCE measurement is a very useful tool to study quantum-critical behavior~\cite{Tokiwa2011RSI,Rost2009Science} as well as various phase transitions~\cite{Aoki2004JPSJ,Kono2018PRB}.

Precise control of the field orientation recently opened the possibility of discovering exotic phenomena in condensed matter.
For example, an unusual first-order superconducting transition has been discovered in Sr$_2$RuO$_4$ 
when a magnetic field is applied very close to (within 2$^\circ$ of) the $ab$ plane~\cite{Yonezawa2013PRL,Yonezawa2014JPSJ,Kittaka2014PRB}.
In addition, tangential features of the wing structure in the $(T, H_b, H_c)$ phase diagram of the Ising ferromagnet URhGe have been found 
by precise application of a magnetic field aligned with the hard $b$ axis within less than 1$^\circ$~\cite{Nakamura2017PRB}.
Here, $H_b$ and $H_c$ are the components of the magnetic field along the $b$ and $c$ axes, respectively.
Moreover, field-angle variation of the quasiparticle density of states, which can be detected from the low-temperature specific heat and thermal conductivity, 
is effective for identifying the gap structure of exotic superconductors~\cite{Sakakibara2016RPP, Matsuda2006JPCM}.
Thus, a number of key experimental facts have been revealed by field-angle-resolved experiments.

In this paper, we propose a new experimental technique to investigate the \textit{field-angle} $\phi_H$ dependence of the entropy, $S(\phi_H)$.
Let us consider that a planar rotating magnetic field $\Vec{H}=H(\cos\phi_H,\sin\phi_H,0)$ is applied to a material  
whose macroscopic magnetization is expressed as $\Vec{M}=M(\sin\theta_m\cos\phi_m,\sin\theta_m\sin\phi_m,\cos\theta_m)$.
The angles $\phi_H$ and $\phi_m$ are the azimuthal angles of $\Vec{H}$ and $\Vec{M}$, respectively, 
and $\theta_m$ is the polar angle of $\Vec{M}$ measured from the direction normal to the field-rotation plane.
Clearly, both $H$ and $\phi_H$ can be tuning parameters of the internal energy $U$ that includes the inner product of $\Vec{H}$ and $\Vec{M}$.
Then, $S$ can be expanded by $T$, $H$, and $\phi_H$ as
\begin{align}
dS &= \frac{d^\prime Q}{T}\notag \\
&=\biggl(\frac{\partial S}{\partial T}\biggl)_{H, \phi_{H}}dT + \biggl(\frac{\partial S}{\partial H}\biggl)_{T, \phi_{H}}dH + \biggl(\frac{\partial S}{\partial \phi_H}\biggl)_{T,H}d\phi_H.
\end{align}
Under adiabatic conditions ($d^\prime Q=0$) and when $\phi_H$ is fixed ($d\phi_H=0$), the following relation can be derived:
\begin{equation}
\biggl(\frac{\partial S}{\partial H}\biggl)_{T,\phi_H}=-\frac{C}{T}\biggl(\frac{\partial T}{\partial H}\biggl)_{S,\phi_H}.
\end{equation}
This relation describes the \textit{conventional} MCE that occurs when the strength of the magnetic field changes.
Similarly, when the magnetic field is rotated with a fixed $H$ ($dH=0$), the \textit{rotational} MCE can be derived as
\begin{equation}
\biggl(\frac{\partial S}{\partial \phi_H}\biggl)_{T,H}=-\frac{C}{T}\biggl(\frac{\partial T}{\partial \phi_H}\biggl)_{S,H}.
\end{equation}
This relation indicates that the field-angle $\phi_H$ dependence of $S$ can be directly investigated by measuring the change in the sample temperature during the field rotation. 

Under quasi-adiabatic conditions, however, careful evaluation of $(\partial S/\partial \phi_H)_{T,H}$ is necessary because 
the heat transfer, $d^\prime Q=-\kappa(T-T_0)dt-d^\prime Q_{\rm loss}$, is finite~\cite{Rost2009Science,Yonezawa2013PRL}.
Here, $\kappa$ is the thermal conductance between the sample and a thermal bath, and $d^\prime Q_{\rm loss}$ is the dissipative loss.
In this case, the rotational MCE can be expressed as
\begin{equation}
\biggl(\frac{\partial S}{\partial \phi_H}\biggl)_{T,H}=-\frac{C}{T}\frac{dT}{d\phi_H}-\frac{\kappa(T-T_0)}{T}\frac{d t}{d\phi_H}-\frac{1}{T}\frac{d^\prime Q_{\rm loss}}{d\phi_H}.
\end{equation}
The main source of the rotational heating $d^\prime Q_{\rm loss}/dt$ is Joule heating by eddy currents, and 
it is proportional to the square of the angular velocity, \textit{i.e.}, $\propto (d\phi_H/dt)^2$.
Then, the last term is proportional to $d\phi_H/dt$, whereas the second term is inversely proportional to it. 
Therefore, it is necessary to tune the rotational speed $d\phi_H/dt$ as well as $\kappa$ and $d^\prime Q_{\rm loss}$ to measure $(\partial S/\partial \phi_H)_{T,H}$ precisely.
Otherwise, averaging $dT/d\phi_H$ values taken under clockwise and anticlockwise field rotation may be effective to cancel out the heat-transfer effect
because $d\phi_H/dt$ changes its sign with reversing field-rotation direction, whereas $(\partial S/\partial \phi_H)_{T,H}$ does not.
 
\begin{figure}
\begin{center}
\includegraphics[width=3.in]{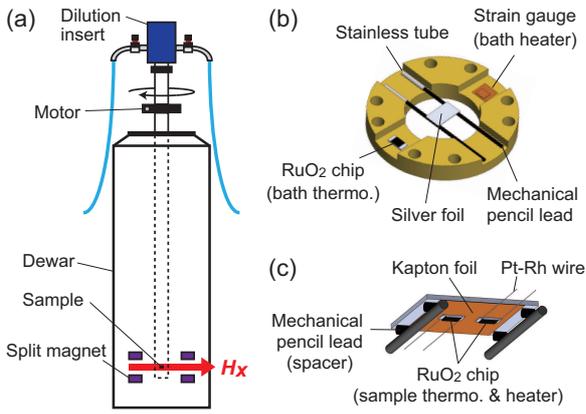}
\caption{
(Color online) Schematic views of (a) the field-orientation control system, (b) a home-made cell to measure the specific heat and MCE, and (c) the addenda of the cell (bottom view).
}
\label{cell}
\end{center}
\end{figure}

To measure the MCE and specific heat in a rotating magnetic field,
we developed the apparatus illustrated in Fig.~\ref{cell}.
To prevent the sample from rotating owing to strong magnetic torque, 
the sample stage, made of a silver foil whose thickness is 0.1 mm, is supported by mechanical pencil leads with a hardness grade of 2H and a diameter of 0.3 mm [Figs.~\ref{cell}(b) and \ref{cell}(c)].
One side of the pencil leads is not fixed but inserted into a stainless tube fixed on the brass frame to avoid stress arising from the difference in thermal expansion between the lead and the frame.
Two thick-film ruthenium oxide chip resistors are attached to the bottom of the addenda with a Kapton foil (7.5 $\muup$m in thickness) for insulation;
they are used as the sample thermometer (KOA, RK73B-1E, 2 k$\Omega$) and heater (KOA, RK73B-1F, 2 k$\Omega$).
The resistance of the sample thermometer is measured using a lock-in amplifier (Stanford Research, SR830) 
with a measurement frequency of 997 Hz and a time constant of 30~ms.
The temperature of a thermal bath is measured by a chip resistor (KOA, RK73B-1E, 2 k$\Omega$) and an ac resistance bridge (Picowatt, AVS-47B), and 
is controlled by a heater (strain gauge) by using a temperature controller (Picowatt, TS-530).
These thermometers were calibrated by a commercial ruthenium oxide temperature sensor (Entropy, RuO$_x$-B).
Fortunately, the anisotropy in the magnetoresistance of the chip resistors is negligibly small~\cite{Deguchi2004RSI}. 
Pt--Rh wires with a diameter of 25 $\muup$m are used as conducting wires 
because NbTi superconducting wires cause heating due to flux depinning during field sweep and rotation.
The magnetic field is generated along the horizontal direction using a split-pair magnet.
As illustrated in Fig.~\ref{cell}(a), a dilution refrigerator is smoothly rotated around the vertical axis 
using a stepper motor (Huber, Goniometer 409) with a gear ratio of 180:1.
Hence, the system makes it possible to control the angle between the magnetic field and a crystalline axis.
The rotational speed is adjusted using a motor driver (Niki Glass, NIKI-PMD-2CH) and a stepping motor controller (Tsuji Electronics, UPM2C-01);
in this study, we set it to $d\phi_H/dt=50$~s/deg.

\begin{figure}
\begin{center}
\includegraphics[width=3.in]{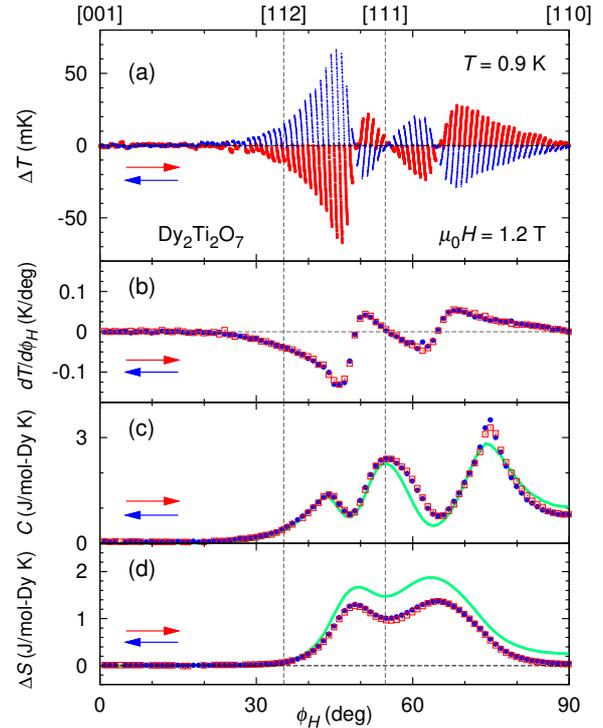}
\caption{
(Color online) (a) Field-angle $\phi_H$ dependence of change in the sample temperature, $\Delta T=T - T_{\rm base}$, during each 1$^\circ$ rotation of the magnetic field (1.2~T) 
within the $(1\bar{1}0)$ plane of $\DTO$ at 0.9~K. 
Here, $T_{\rm base}$ is the base temperature of the sample, and $\phi_H$ is the field angle measured from the $[001]$ axis in the $(1\bar{1}0)$ plane.
Field-angle variation of (b) the initial slope, $dT/d\phi_H$, just after the field rotation began, 
(c) the specific heat $C$, and (d) the entropy $\Delta S$ measured from $S(\phi_H=0^\circ)$.
Solid lines in (c) and (d) are results of Monte Carlo simulations at 0.9~K and 1.2~T~\cite{Kittaka2018b}.
}
\label{MCE}
\end{center}
\end{figure}

To test the apparatus,
we chose the spin-ice compound Dy$_2$Ti$_2$O$_7$ \cite{Bramwell01,Castelnovo08,Kadowaki09} as a benchmark 
and measured the specific heat and MCE under a rotating magnetic field within the cubic $(1\bar{1}0)$ plane.
Here, $\phi_H$ is defined as the in-plane field angle measured from the [001] axis.
The single-crystal sample of Dy$_{2+x}$Ti$_{2-x}$O$_{7+y}$ ($x=0.002 \pm 0.002$) used in this study was cut from
a crystal rod grown by the floating-zone method described in Ref. \ref{Kadowaki2018}. 
The sample dimensions are roughly $1.8 \times 1.4 \times 0.14$~mm$^3$.
The shortest dimension is along the $[1\bar{1}0]$ axis so that the demagnetization factor becomes small and less anisotropic 
during the field rotation.

Figure~\ref{MCE} presents an example set of a field-angle-resolved measurement of the entropy of Dy$_2$Ti$_2$O$_7$.
In this measurement, 
the temperature of the bath thermometer was controlled so that the sample temperature became stable at 0.9~K when $\Vec{H}$ was fixed.
Then, the applied magnetic field of 1.2~T was rotated by 1$^\circ$ within the $(1\bar{1}0)$ plane, and 
the sample temperature was recorded during the rotation [Fig.~\ref{MCE}(a)].
The initial slope of $T(\phi_H)$ just after the field rotation began, $dT/d\phi_H$, is plotted in Fig.~\ref{MCE}(b).
For each angle $\phi_H$, after the sample temperature became stable again, 
the specific heat was measured by the quasi-adiabatic heat-pulse method [Fig.~\ref{MCE}(c)].
After taking the specific-heat data, we again waited until the sample temperature became stable 
and then moved to the next data point by rotating $\Vec{H}$ by 1$^\circ$.

To estimate the heat-transfer effect, the above measurements were performed
under both clockwise and anticlockwise field rotation, as represented in Fig.~\ref{MCE} by different colors.
As clearly seen in Fig.~\ref{MCE}(a), the sample temperature is switched between increasing and decreasing by changing the direction of the field rotation.
This result demonstrates that the observed behavior indeed arises from the rotational MCE.
The initial slopes, $dT/d\phi_H$, evaluated from clockwise and anticlockwise rotation measurements [Fig.~\ref{MCE}(b)] coincide well with each other, 
indicating that the heat-transfer effect is negligible under the present condition.

\begin{figure}
\begin{center}
\includegraphics[width=3.2in]{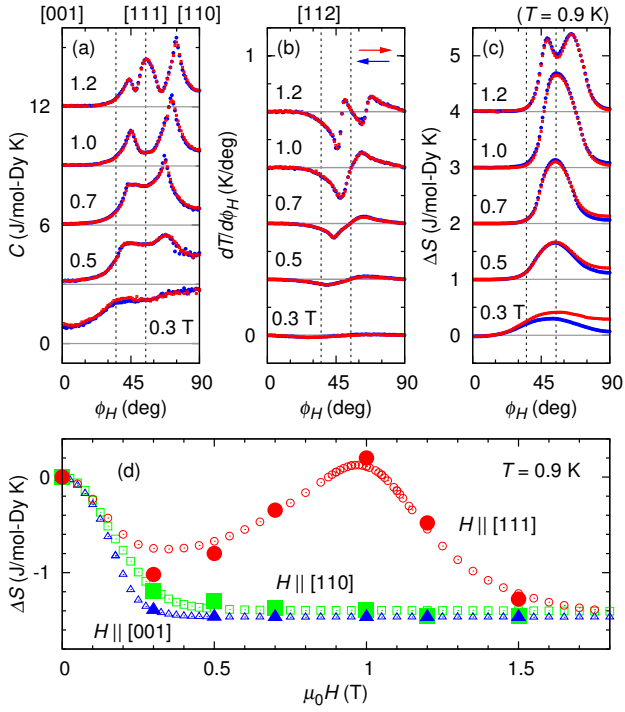}
\caption{
(Color online) Field-angle $\phi_H$ dependence of (a) $C$, (b) $dT/d\phi_H$, and (c) $\Delta S$ of $\DTO$ at 0.9 K in various magnetic fields.
Each set of data is shifted vertically by (a) 3~J/(mol-Dy K), (b) 0.2~K/deg, and (c) 1~J/(mol-Dy K) for clarity.
(d) Relative change in entropy as a function of the magnetic field strength, $\Delta S(H)=S(H)-S(H=0)$, at 0.9~K for $H \parallel$ [001] (triangles), [111] (circles), and [110] (squares) obtained from conventional (open) and rotational (closed) MCE measurements.
The data obtained under rotation, $\Delta S(\phi_H)$ at $0^\circ$, $54.7^\circ$, and $90^\circ$, are plotted so that $\Delta S(\phi_H=0^\circ)$ is equal to $\Delta S(H)$ for $H \parallel [001]$.
}
\label{MCE09}
\end{center}
\end{figure}

Figure~\ref{MCE}(d) shows the field-angle dependence of the entropy 
\begin{equation}
\Delta S(\phi_H)=-\int_0^{\phi_H}\frac{C}{T}\frac{dT}{d\phi_H}d\phi_H
\end{equation}
obtained using the data in Figs. \ref{MCE}(b) and \ref{MCE}(c).
Here, we define $\Delta S(\phi_H)$ as the entropy change relative to $S(\phi_H=0^\circ)$. 
To check the reliability of the present rotational MCE measurements, 
Monte Carlo simulations based on the dipolar spin-ice model were conducted~\cite{Kittaka2018b}.
The calculated $C(\phi_H)$ and $\Delta S(\phi_H)$ at 0.9~K and 1.2~T are shown in Figs. \ref{MCE}(c) and \ref{MCE}(d) as solid lines, respectively.
The measured $C(\phi_H)$ and $\Delta S(\phi_H)$ agree with the simulated curves at the same level as those shown in Figs.~1(c) and 1(d) of Ref.~\ref{Tabata2006PRL}. 

The field variation of $\Delta S(\phi_H)$ at 0.9~K for both clockwise and anticlockwise field rotation is shown in Fig. \ref{MCE09}(c), 
which was obtained using the data for $C(\phi_H)$ [Fig. \ref{MCE09}(a)] and $dT/d\phi_H$ [Fig. \ref{MCE09}(b)].
In Fig. \ref{MCE09}(d), the relative change in $S(H)$ with respect to $S(H=0)$ is plotted as open symbols for $H \parallel [001]$, [111], and [110],
which are obtained from conventional MCE measurements using the same setup; 
the present data for $S(H)$ are consistent with a previous report~\cite{Hiroi2003JPSJ,Hiroi2003JPSJ-2}.
The values of $\Delta S(\phi_H)$ at $\phi=0^\circ$, $54.7^\circ$, and $90^\circ$ are also plotted in Fig. \ref{MCE09}(d) as closed symbols 
so that $\Delta S(\phi_H=0^\circ)$ is equal to $\Delta S(H)$ at $\phi_H=0^\circ$. 
Good agreement can be seen except at 0.3~T.
This inconsistency at 0.3~T is probably due to spin freezing in the low-field state of Dy$_2$Ti$_2$O$_7$, which produces a strong non-equilibrium state~\cite{Pomaranski2013NatPhys,Paulsen2014NatPhys}.
Indeed, a weak but clear hysteresis is observed in $\Delta S(\phi_H)$ at 0.3~T, whereas it is negligible at other fields.

\begin{figure}
\begin{center}
\includegraphics[width=3.in]{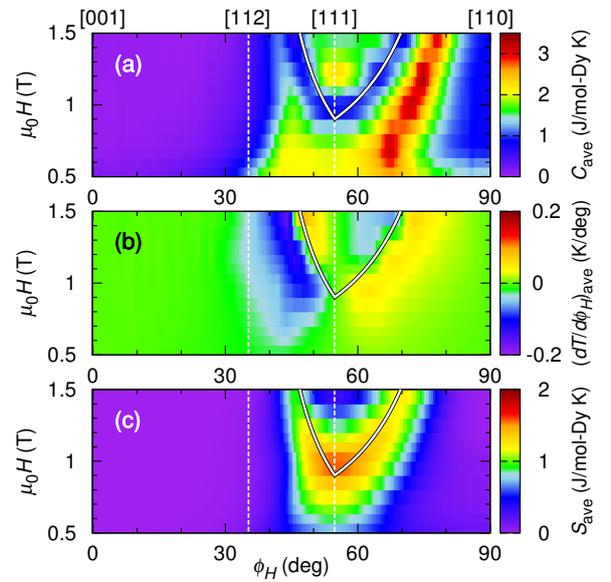}
\caption{
(Color online) Contour plots of (a) $C(H,\phi_H)$, (b) $dT/d\phi_H(H,\phi_H)$, and (c) $S(H,\phi_H)$ of $\DTO$ at 0.9~K obtained by averaging the clockwise and anticlockwise field-rotation data.
In (c), $\Delta S(\phi_H)$ values measured at different fields are connected using the conventional MCE data in Fig.~\ref{MCE09}(d),
and the absolute value of $S_{\rm ave}(H,\phi_H=0^\circ)$ is assumed to be zero at $\mu_0H= 1.5$~T.
Solid lines represent the critical field $H_{\rm c}$ calculated on the basis of a nearest-neighbor spin-ice model with $J_{\rm eff}=1.01$~K~\cite{Sato2007JPCM}.
}
\label{map09}
\end{center}
\end{figure}

By combining the data for $\Delta S(H)$ and $\Delta S(\phi_H)$, 
it is possible to construct a contour map of $S_{\rm ave}(\phi_H,H)$ [Fig.~\ref{map09}(c)].
Here, $S_{\rm ave}$ is the average of $\Delta S(\phi_H)$ under clockwise and anticlockwise rotation with an offset;
the offset is determined so that $S_{\rm ave}=0$ at $\phi_H=0^\circ$ and $\mu_0H=1.5$~T, because a non-degenerate 2-in 2-out state is realized at high fields for $H \parallel [001]$~\cite{Hiroi2003JPSJ}.
Likewise, $(\phi_H,H)$ contour maps of the averaged $C$ and $dT/d\phi_H$, \textit{i.e.}, $C_{\rm ave}$ and $(dT/d\phi_H)_{\rm ave}$, are also presented in Figs.~\ref{map09}(a) and \ref{map09}(b), respectively.
It has been found that the high-entropy area spreads in a relatively wide $\phi_H$ region at 0.9~K, 
although it is expected that, in the low-field region below $\sim 0.9$~T, a triply degenerate 2-in 2-out state in $H \parallel [111]$ (the kagome-ice state~\cite{Matsuhira2002JPCM,Higashinaka2003PRB,Udagawa2002JPSJ})
is transformed into a macroscopically non-degenerate 2-in 2-out state 
by slightly tilting the magnetic field~\cite{Moessner2003PRB,Fennell2007NatPhys}.

In the high-field region for $H \parallel [111]$, a spin-flip transition from the highly degenerate kagome-ice state to a fully ordered 3-in 1-out (1-in 3-out) state occurs owing to the Zeeman effect~\cite{Sakakibara2003PRL}.
The solid lines in Figs.~\ref{map09}(a)--\ref{map09}(c) represent the $\phi_H$ dependence of the spin-flip critical field $H_{\rm c}$ derived from the nearest-neighbor spin-ice model~\cite{Sato2007JPCM}:
\begin{align}
H_{\rm c}&=\frac{6J_{\rm eff}}{\cos\phi-2\sqrt{2}\sin\phi}\ \ (\phi \ge 0),\\
H_{\rm c}&=\frac{6J_{\rm eff}}{\cos\phi+\sqrt{2}\sin\phi}\ \ (\phi \le 0).
\end{align}
Here, $\phi$ ($=\phi_{111}-\phi_H$) is the angle between $H$ and the $[111]$ direction ($\phi_{111}\simeq 54.7^\circ$), and 
$J_{\rm eff}$ ($=1.01$~K for $\DTO$) represents an effective nearest-neighbor interaction consisting of an antiferromagnetic exchange and a ferromagnetic dipolar interaction.
The $S_{\rm ave}(\phi_H,H)$ data successfully capture the $H_{\rm c}(\phi_H)$ line, a level crossing from the 2-in 2-out to the 3-in 1-out (1-in 3-out) state.
In contrast, $C_{\rm ave}(\phi_H,H)$ in Fig.~\ref{map09}(a) shows a rather complicated map; 
it is hard to extract information on the spin configuration and the $H_{\rm c}(\phi_H)$ line.

\begin{figure}
\begin{center}
\includegraphics[width=3.2in]{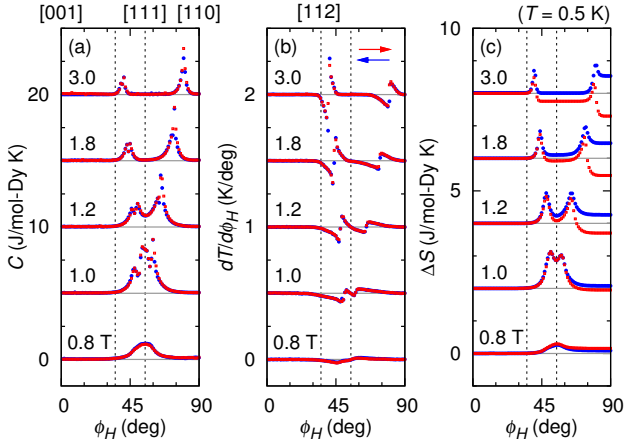}
\caption{
(Color online) Field-angle $\phi_H$ dependence of (a) $C$, (b) $dT/d\phi_H$, and (c) $\Delta S$ of $\DTO$ at 0.5~K in various magnetic fields.
Each set of data is shifted vertically by (a) 5~J/(mol-Dy K), (b) 0.5~K/deg, and (c) 2~J/(mol-Dy K) for clarity.
}
\label{MCE05}
\end{center}
\end{figure}

\begin{figure}
\begin{center}
\includegraphics[width=3.in]{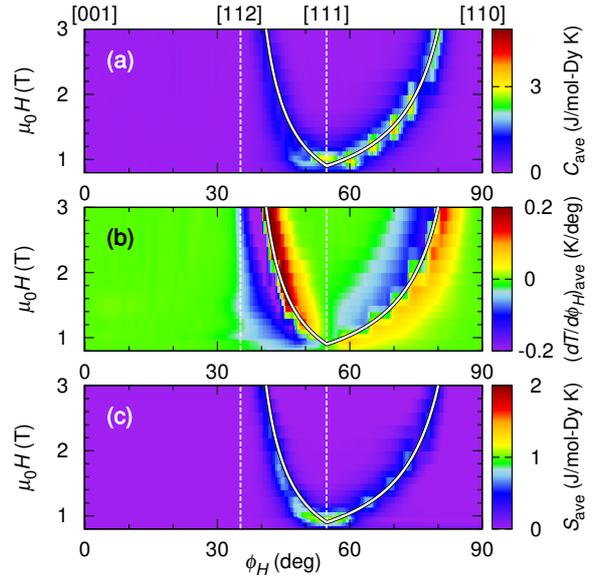}
\caption{
(Color online) Contour plots of (a) $C(H,\phi_H)$, (b) $dT/d\phi_H(H,\phi_H)$, and (c) $\Delta S(H,\phi_H)$ of $\DTO$ at 0.5~K obtained by averaging their clockwise and anticlockwise field-rotation data.
In (c), we assume $S_{\rm ave}(H,\phi_H)=0$ at $\phi_H=0^\circ$ and $\mu_0H\ge 0.8$~T.
Solid lines are the same as those in Fig.~\ref{map09}.
}
\label{map05}
\end{center}
\end{figure}

As the temperature decreases to 0.5~K, the $H_{\rm c}$ anomalies in $C(\phi_H)$, $dT/d\phi_H$, and $\Delta S(\phi_H)$ become sharper and clearer, as shown in Figs.~\ref{MCE05}(a)--\ref{MCE05}(c),
although the sample is far from equilibrium in the low-field region below $\sim 0.8$~T~\cite{Kittaka2018}.
Figures \ref{map05}(a)--\ref{map05}(c) show the $(H,\phi_H)$ contour maps of $C_{\rm ave}$, $(dT/d\phi_H)_{\rm ave}$, and $S_{\rm ave}$ at 0.5~K, respectively;
the anomalies are concentrated in the region close to the $H_{\rm c}(\phi_H)$ line.
This sharpness of the anomalies is due to a first-order spin-flip transition in Dy$_2$Ti$_2$O$_7$ that occurs below 0.3~K~\cite{Sakakibara2003PRL,Sato2006JPCM}.
Indeed, a slight, weak hysteresis is observed around the peaks in $C(\phi_H)$ above 1.2~T, confirming the first-order nature of the spin-flip transition.
Note that a large hysteresis is observed in $\Delta S(\phi_H)$ because it is obtained by integration of $C/T$ multiplied by $dT/d\phi_H$.
Thus, $\Delta S(\phi_H)$ is quite sensitive to the first-order nature of the phase transitions as well as the presence of non-equilibrium states.

Finally, let us discuss the thermodynamic relation of $S(H,\phi_H)$.
According to the Maxwell relation, we obtain
\begin{align}
\biggl(\frac{\partial S}{\partial H}\biggl)_{T,\phi_H}&=\biggl(\frac{\partial M_\parallel}{\partial T}\biggl)_{H,\phi_H}\\
\biggl(\frac{\partial S}{\partial \phi_H}\biggl)_{T,H}&=\biggl(\frac{\partial \tau_{\phi}}{\partial T}\biggl)_{H,\phi_H},
\end{align}
where $M_\parallel$ [$=M\cos(\phi_m-\phi_H)\sin\theta_m$] is the component of $\Vec{M}$ parallel to the field direction, and $\tau_\phi$ [$=MH\sin(\phi_m-\phi_H)\sin\theta_m$] is the in-plane magnetic torque.
Therefore, $(\partial S/\partial \phi_H)_{T,H}=0$ represents $\tau_\phi=0$, or temperature-independent $\tau_\phi$ under the measurement conditions.
By taking advantage of this relation, symmetric crystalline axes enforcing $\tau_\phi=0$ can be precisely determined;
\textit{e.g.}, as demonstrated in Figs.~\ref{map09}(b) and \ref{map05}(b), 
$(dT/d\phi_H)_{\rm ave}$ [$\propto (\partial S/ \partial \phi_H)_{T,H}$] becomes zero at any field and any temperature when the magnetic field is aligned with the [001], [111], or [110] direction (in the equilibrium condition).
Otherwise, $(dT/d\phi_H)_{\rm ave}$ also becomes zero when $S(\phi_H)$ has a local maximum (or minimum),
corresponding to the spin-flip $H_{\rm c}(\phi_H)$ line.
Thus, rotational MCE measurements offer a new and strong approach to determining the angle dependence of various phase transitions as well as the orientation of the crystalline axes.

In summary, we developed a new technique to measure the field-angle dependence of the low-temperature entropy with high resolution and high speed using the rotational MCE.
Using this technique, we succeeded in detecting the well-known spin-flip transition in the spin-ice compound $\DTO$ under a rotating magnetic field within the $(1\bar{1}0)$ plane.
Because the entropy is a thermodynamic quantity, the experimental data for $S(T,H,\phi_H)$ can be compared with calculated results by solving a model Hamiltonian.
Therefore, this technique has the potential for development as a powerful tool to study field-angle-dependent exotic phenomena in a wide variety of research fields.

\acknowledgments
The computation in this work was done using the facilities of the Supercomputer Center, the Institute for Solid State Physics, the University of Tokyo.
This work was supported by a Grants-in-Aid for Scientific Research on Innovative Areas ``J-Physics'' (15H05883, 18H04306)
from MEXT and KAKENHI (18H01161, 18H01164) from JSPS.


\end{document}